\documentclass{aastex631}  
\pdfoutput=1
\usepackage{xcolor}
\usepackage{enumitem}
\begin{document}
\graphicspath{{./}{figures/}}

%% ----------------------------------------------------------

\author[0000-0002-6130-7829]{Shaonwita Pal}
\affiliation{Center of Excellence in Space Sciences India,\\ Indian Institute of Science Education and Research Kolkata, Mohanpur 741246, West Bengal, India}

\author[0000-0002-4409-7284]{Prantika Bhowmik}
\affiliation{Department of Mathematical Sciences, Durham University, Durham, DH1 3LE, UK}

\author[0000-0003-1753-8002]{Sushant S. Mahajan}
\affiliation{W.W. Hansen Experimental Physics Laboratory, Stanford University, Stanford, CA 94305, USA}

\author[0000-0001-5205-2302]{Dibyendu Nandy}
\altaffiliation{dnandi@iiserkol.ac.in}
\affiliation{Center of Excellence in Space Sciences India,\\ Indian Institute of Science Education and Research Kolkata, Mohanpur 741246, West Bengal, India}
\affiliation{Department of Physical Sciences,\\ Indian Institute of Science Education and Research Kolkata, Mohanpur 741246, West Bengal, India}

%%---------------------------------------------------------------

\title{Impact of Anomalous Active Regions on the Large-scale Magnetic Field of the Sun}

\begin{abstract}
One of the major sources of perturbation in the solar cycle amplitude is believed to be the emergence of anomalous active regions which do not obey Hale's polarity law and Joy's law of tilt angles. Anomalous regions containing high magnetic flux that disproportionately impact the polar field are sometimes referred to as ``rogue regions". In this study -- utilizing a surface flux transport model -- we analyze the large-scale dipole moment build-up due to the emergence of anomalous active regions on the solar surface. Although these active regions comprise a small fraction of the total sunspot number, they can substantially influence the magnetic dipole moment build-up and subsequent solar cycle amplitude. Our numerical simulations demonstrate that the impact of ``Anti-Joy'' regions on the solar cycle is similar to those of ``Anti-Hale'' regions. We also find that the emergence time, emergence latitude, relative number and flux distribution of anomalous regions influence the large-scale magnetic field dynamics in diverse ways. We establish that the results of our numerical study are consistent with the algebraic (analytic) approach to explaining the Sun's dipole moment evolution. Our results are relevant for understanding how anomalous active regions modulate the Sun's large-scale dipole moment build-up and its reversal timing within the framework of the Babcock-Leighton dynamo mechanism -- now believed to be the primary source of solar cycle variations.
\end{abstract}

\section{Introduction} \label{sec:intro}

Active regions emerging on the solar surface have long been considered the primary element through which the solar cycle manifests itself \citep{Schwabe1844}. Short and long-term evolution of the magnetic field associated with active regions drives solar magnetic events such as flares, coronal mass ejections, energetic particle releases, etc., besides controlling the Sun's total electromagnetic output. These activities subsequently affect the Earth and its vicinity, specifically satellite operations, telecommunications, and contribute to various aspects of space weather \citep{Nandy2004, Nandy2007,Schriver2015}. Since surface magnetic field distribution and its evolution play crucial roles in governing the short and long-term changes in space weather and space climate, respectively, it is vital to study the large-scale surface magnetic field during a solar cycle as well as its cycle to cycle irregularities \citep{Bhowmik2018, Petrovay2020, Nandy2021, Bhowmik2023}. The magnetic cycle of the Sun can be explained using the Babcock-Leighton (B-L) Solar Dynamo theory \citep{Leighton1964, Piddington1972,  Wang1991, Hazra2016, Charbonneau2020, Fan2021, Hazra2023}. At the solar activity minimum, the global magnetic field is primarily dominated by the poloidal field component, which is predominantly dipolar. The solar differential rotation stretches this poloidal field longitudinally to generate the toroidal field component in the tachocline region. Amplified toroidal flux ropes eventually encounter magnetic buoyancy in the convection zone and emerge primarily as Bipolar Magnetic Regions (BMRs) \citep{Nandy2002,Gilman2018,Fan2021}. Near the Sun's equator, the leading sunspot polarities of two opposite hemispheres annihilate each other, and the remaining trailing spot (mainly of uni-polar magnetic fields) drifts towards the respective poles via large-scale meridional flow and diffusion. During the first half of the solar cycle, many such trailing polarities reaching the polar region eventually cancel the existing large-scale polar field and create a new one with the opposite sign. This reversal of the global dipolar field happens during the solar cycle maximum, and the new polar field keeps growing till the end of the cycle. This whole process of redistribution of active region-associated magnetic flux on the solar surface is known as the B-L mechanism \citep{Babcock1961,Leighton1964,Mackay2012,jiang2014b}. One of the existing hypotheses explaining the dynamo cycle involves meridional circulation advecting the poloidal field down to the tachocline region from the solar photosphere followed by the generation and enhancement of toroidal field via Parker’s $\Omega$-effect \citep{Parker1955} - thus completing the full solar cycle. Therefore, the B-L mechanism is the primary means for evolving the polar field and also a crucial component controlling variability in the amplitude of sunspot cycles over decadal to century-scales \citep{Nandy2001, Dasi2010, Munoz2010, Bhowmik2018,Kumar2019, DASH2020AGU, Nandy2023arxiv}.

Active regions associated with a solar cycle have their own characteristics and largely follow specific patterns. They emerge closer to the equator as the cycle progresses. This latitudinal variation of active regions in both hemispheres primarily follows Sp$\ddot{o}$rer's Law \citep{Carrington1858}. Hale's polarity law \citep{Hale1919,Hale1925} determines the sign of magnetic polarities (positive/negative) of active regions also known as Bipolar Magnetic Regions (BMRs) in each hemisphere. According to Hale's law, the relative order of magnetic polarities of the leading and trailing spots remains the same in a particular hemisphere, but it's the opposite between the northern and southern hemispheres. To elaborate, if, during a particular cycle, the leading spots of BMRs emerging in the northern hemisphere have positive polarity, the trailing spots will be of negative polarity. The pattern would be opposite for the southern hemispheric BMRs with negative leading spots and positive trailing spots. This polarity order also reverses every solar cycle. Observations suggest that for most sunspot groups, their leading and following spots are seated close to the equator and the pole, respectively. Thus, the magnetic axis joining the centers of the two spots (leader and follower) has a weak statistical tendency to have a slight positive (negative) tilt angle with the axis parallel to the solar equator in the northern (southern) hemisphere. The amplitude of tilt angles generally increases with increasing latitude. This relation between active regions' emergence latitudes and tilts is well known as Joy's tilt Law \citep{Hale1919}. Coriolis force, responsible for twisting the toroidal flux tube during the time of emergence is believed to cause this tilt angle \citep{Choudhuri1987,Fisher1995, Kleeorin2020}.

Sp$\ddot{o}$rer's law, Joy's tilt law, and Hale's polarity law - these three well-observed laws that describe the location and orientation of sunspot pairs on the solar surface, play a key role in understanding the solar magnetic cycle and its variations through B-L mechanism. The emerging sunspots which do not obey these laws can significantly impact the long-term behaviour of the large-scale solar magnetic field. We have termed these spots `anomalous regions'. Rogue region is also one category of anomalous regions whose flux and tilt angle is very high \citep{Nagy2017, Nagy2019}. In short, `anomalous' classification introduced in this manuscript corresponds to all categories of Anti-hale and Anti-joy regions (or combinations of both) irrespective of the amount of flux and degree of tilt angle. We investigate the strength of impact on the solar cycle for these different categories of anomalous regions due to their diversity of tilt characteristics, flux content and spatio-temporal distribution.

There are several theories on the generation of such anomalous regions. Anti-Joy regions (with tilt angles opposite from what should be according to their latitudinal positions) emerge due to two reasons primarily: (1) randomness in Coriolis force \citep{Schmidt1968}, (2) convective buffeting of flux tubes \citep{Weber2013}. For the Anti-Hale regions (regions with opposite polarity orientation), studies suggest that they can be formed due to (1) kink instability in highly twisted magnetic flux tube \citep{Nandy2006, Knizhnik2018}, (2) oppositely oriented toroidal flux-tubes within the convection zone \citep{Stenflo2012}, (3) transport of Hale regions from opposite hemisphere \citep{McClintock2014}, and (4) small-scale dynamo action creating sunspots at the end of the cycle \citep{Sokoloff2015}. 

Past studies suggest that tilt quenching and variation in the meridional circulation are vital factors causing cycle irregularities \citep{Wang2002ApJ,Dasi2010,Upton2014}. Additionally, the scattering in active regions' tilt angles and Anti-Joy regions produce significant changes in large-scale polar field build-up and open magnetic flux dynamics during a solar cycle \citep{Cameron2010,Jiang2014a}. Besides, the appearance of a single large Anti-Joy and Anti-Hale region can have a substantial effect on cycle amplitude, which was argued by past studies \citep{Yeates2015, Jiang2015} as a probable cause of weaker peaks during solar cycles 23 and 24. A large rogue region emerging at different cycle phases and latitudes with varying fluxes also affects the subsequent sunspot cycle peak activity \citep{Nagy2017, Nagy2019}. Previous observation-based studies claimed that the Anti-Hale or Anti-Joy spots appearing on the solar surface constitute a certain percentage of the total number of sunspots in a cycle, varying from 3$\%$ to 10$\%$ \citep{McClintock2014,Li2018, Zhukova2020, Andres2021}. Thus, it is essential to explore the effect of a group of these anomalous regions on the large-scale surface magnetic field distribution and the overall polar magnetic field variability - which we investigate in this present work. Our detailed quantitative analyses are based on multiple numerical simulations using a data-driven surface flux transport model \citep{Bhowmik2018}. In these simulations, we consider three classes of anomalous regions based on the orientations and polarities of the two spots within the active regions and study the consequences individually. We also investigate how their effects vary depending on the following factors: their latitudinal position, emergence timing relative to the cycle phase, associated magnetic flux and abundance compared to remaining standard Hale-Joy regions. 

The presented work is assembled as follows: in Section \ref{sec:model}, we describe the computational model used in our study. Section \ref{sec:results} summarizes the results we obtain from our simulations. In section \ref{theory}, we validate the results using algebraic approximation and finally, we present our conclusions with relevant discussion in Section \ref{sec:conclusion}.

\section{Simulation Set-up} \label{sec:model}

\subsection{Surface Flux Transport Model}
\label{sec2.2}
 Surface Flux Transport (SFT) models are utilized to demonstrate the time evolution of the large-scale photospheric magnetic field distribution according to the Babcock-Leighton mechanism. This model includes the effects of supergranular turbulent diffusion ($\eta$) and advective transport through large-scale plasma flows like differential rotation [$\omega$($\theta$)] \& meridional circulation [$\mathrm{v}$($\theta$)] and particularly solves the radial part of the magnetic induction equation which is given by, 

\begin{equation}
\frac{\partial B_{r}}{\partial t} = - \omega(\theta)\frac{\partial B_{r}}{\partial \phi} - \frac{1}{\mathrm{R_\odot}\,\sin\theta}\frac{\partial }
{\partial \theta}\Big(v(\theta)B_{r}\sin\theta \Big) + \frac{\eta}{\mathrm{R_\odot}^{2}}\Bigg[\frac{1}{\sin\theta}\frac{\partial }{\partial \theta}\left(\sin\theta \frac{\partial B_{r}}{\partial \theta} \right)
+ \frac{1}{\sin^{2}\theta}\frac{ \partial^{2} B_{r}}{\partial \phi^{2}}\Bigg] + S(\theta,\phi,t).
\label{eq2}
\end{equation}

\noindent Where B$_{r}$ denotes the radial component of the magnetic field, $\theta$ and $\phi$ are colatitude and longitude, respectively. We choose a constant turbulent magnetic diffusivity of $\eta$ = 250 km$^{2 }$s$^{-1}$ for our model (the same value was used in calibrated century-scale observational data-driven SFT simulations by \citeauthor{Bhowmik2018}, \citeyear{Bhowmik2018}). The source term, S($\theta$,$\phi$,t), mimics the emergence of new active regions on the solar surface. $R_\odot$ is the solar radius. Equation (\ref{eq2}) is solved in a domain, $\theta: 0$ to $\pi$ and $\phi: 0$ to $2\pi$ radians, i.e., covering the whole photosphere. The same SFT code based on spherical harmonics expansions (with degrees, $l$: 1 to 63) has been used in many past studies \citep{Nandy2018,Bhowmik2018,Bhowmik2019}.

To include the observed large-scale axisymmetric plasma flow in the toroidal direction, known as differential rotation, $\omega$($\theta$), we rely on the following mathematical expression, provided by \cite{Snodgrass1983},
\begin{equation}
\omega(\theta) = 13.38 - 2.30\:cos^{2}\theta - 1.62\: cos^{4}\theta\,\,\: \textrm{degrees ${\mathrm{day}}^{-1}$.}
\label{eq3}
\end{equation}

\noindent This profile is in accordance with the helioseismic observation \citep{Durney1974, Schou1998}. In addition, there is another observed weak large-scale plasma flow named meridional circulation \citep{Hathaway1993,1993SoPh..147..207K,Mahajan2021,Shravan2022}, which helps in transporting magnetized plasma from the equatorial region to the poles in respective hemispheres. In our simulation set-up, this flow is bounded within $0^{\circ}$ to $\pm$75$^{\circ}$ in each hemisphere, attains the peak amplitude in the mid-latitudes and becomes zero at the equator. We choose a time-independent meridional velocity profile similar to \cite{vanBallegooijen1998},  
\begin{eqnarray}
 v(\lambda) &=
    - v_{0} \hspace{0.1cm \sin( \pi \lambda/ \lambda_0)} & \hspace{5 mm} \text{if } |\lambda | < \lambda_0  \\
    v(\lambda) &=
    0    & \hspace{5 mm} \text{if } |\lambda | > \lambda_0  \\
\label{eq4}
\end{eqnarray}\\

\noindent Where, $\lambda$ is latitude and $\lambda_{0}$ is the limiting latitude ($\lambda_{0}$ = 75$^{\circ}$) beyond which meridional flow vanishes. The amplitude of the meridional flow (${v_0}$) is set at 15 m/s. We initialize our simulation with a dipolar configuration of surface magnetic field concentrated primarily near the 20$^{\circ}$ broad polar cap region ($\pm$70$^{\circ}$ to $\pm$90$^{\circ}$) in both hemispheres. The unsigned strength of the initial polar flux in one hemisphere is nearly about 2 x $10^{22}$ Mx which is consistent with observations. 

%\Sushant{(We might need a reference here for 4.27 Gauss (look in the Webber 2013 paper). Or else, we can say a polar flux of $\sim 10^{22}$ Mx. Please quote the exact polar flux used in the simulation by measuring it.)}

\subsection{Synthetic Sunspot Cycle Profile}
\label{sec2.2}
Modeling the source term [$S(\theta,\phi,t)$] in equation (\ref{eq2}) in SFT simulations requires information on emergence latitude, longitude, tilt angle and area of each region which can be constrained by observed properties of active regions. For our study, we use the synthetic solar cycle data which is prepared utilizing mathematical relations based on general properties of observed solar cycles and their associated active regions. First, we consider a time-dependent Gaussian curve to reproduce the sunspot number time series following \cite{Hathaway2010}. The time-latitude distribution band of active regions is inspired by \cite{Jiang2011}. We use the following square root relation, $\alpha$ = C $\sqrt{|\lambda|}$ to incorporate Joy's tilt law in our synthetic data set. Here, $\alpha$ is the tilt angle and $\lambda$ is the latitudinal position of the centroid of the BMR. The tilt angle of BMRs can be constrained with the variation of cycle strength and the localized inflows, incorporated in the constant factor ``C" \citep{Jiang2014a}. We consider a uniform distribution of active regions over the longitudinal domain of the solar photosphere. 

Since large sunspots rarely emerge during the initial and declining phase of a cycle, it is preferable to use a polynomial distribution fitted to sunspot observations to model their area distribution \citep{Jiang2011,Andres2015}. In order to prescribe the associated magnetic flux, we follow the relationship: $\Phi$(A) = 7.0 $\times$ 10$^{19}$ A (Maxwell), where A represents the area of the whole active region in micro-hemispheres \citep{Dikpati2006}. We assume that all active regions are typical $\beta$-spots having equal flux in two polarities. We model each BMR with a Gaussian profile prescribed by \cite{vanBallegooijen1998}.

\subsection{Preparation of Anomalous Active Regions}
\label{sec2.3}
\begin{figure}
\centering
\includegraphics[width=0.7\textwidth,clip=]{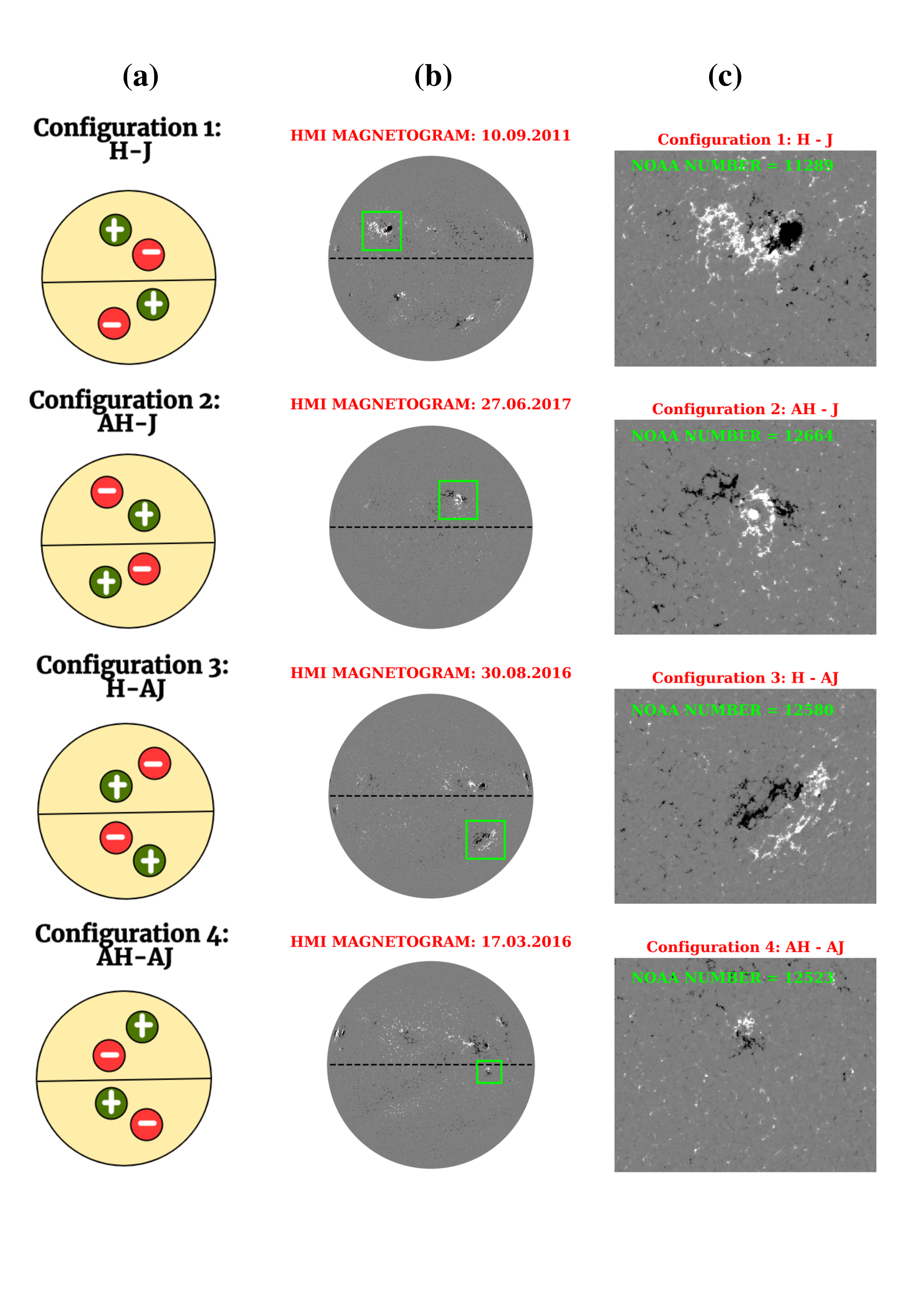}
\caption{\textit{Panel (a):} cartoon of bipolar magnetic region (BMR) with different orientations and polarities in each hemisphere. Here BMR of configuration 1 follows both Hale's and Joy's law (H-J region); configuration 2 represents regions which follow Joy's law but violate Hale's polarity law (AH-J region); configuration 3 does not follow Joy's tilt law but obeys Hale's law (H-AJ region) and lastly, configuration 4 disobeys both Hale's and Joy's law (AH-AJ region). \textit{Panel (b):} surface magnetic field distribution observed in line-of-sight magnetogram from the Helioseismic Magnetic Imager (HMI) instrument on board the Solar Dynamics Observatory (SDO). It denotes four full disk HMI magnetograms observed in Solar Cycle 24. \textit{Panel (c):} cut outs of the green box (in panel b) that serve as examples of the four configurations (H-J, AH-J, H-AJ and AH-AJ regions).}\label{fig:1}
\end{figure}

As discussed earlier, a few emerging active regions may have different orientations and polarities in contrast to the standard Joy's tilt law and Hale's polarity law. We categorize four possible combinations in which three are considered (see the last three cases in Figure~\ref{fig:1}a) as candidates of anomalous regions. 
\begin{itemize}
\item Configuration 1 (H-J): the standard Hale-Joy BMRs (non-anomalous) with negative leading polarity and positive following  polarity in the northern hemisphere and the opposite pattern in the southern hemisphere (see first case in Figure~\ref{fig:1}a). These strictly follow Joy's tilt law and are the ``ideal" spots (denoted by H-J hereafter).

\item Configuration 2 (AH-J): BMR that does not follow Hale's polarity law but follow Joy's tilt law. These configurations are also known as Anti Hale-Joy BMRs (AH-J hereafter).

\item Configuration 3 (H-AJ): BMR that follows Hale's polarity law but have an opposite tilt, unlike H-J, thus, violating Joy's tilt law. As an example, consider the expected positive tilt angle of a BMR in the northern hemisphere to be negative instead. Such a BMR is categorized as a Hale-Anti Joy BMR (H-AJ hereafter).

\item Configuration 4 (AH-AJ): this is the rarest one where the BMR neither follows Hale's polarity law nor Joy's tilt law. It is then an Anti Hale-Anti Joy BMR (AH-AJ hereafter).
\end{itemize}

The middle and last columns of Figure~\ref{fig:1} show observational evidence of the above configurations of active regions appearing during Solar Cycle 24. 

\begin{figure}[h] 
 \centerline{\includegraphics[width=0.8\textwidth,clip=]{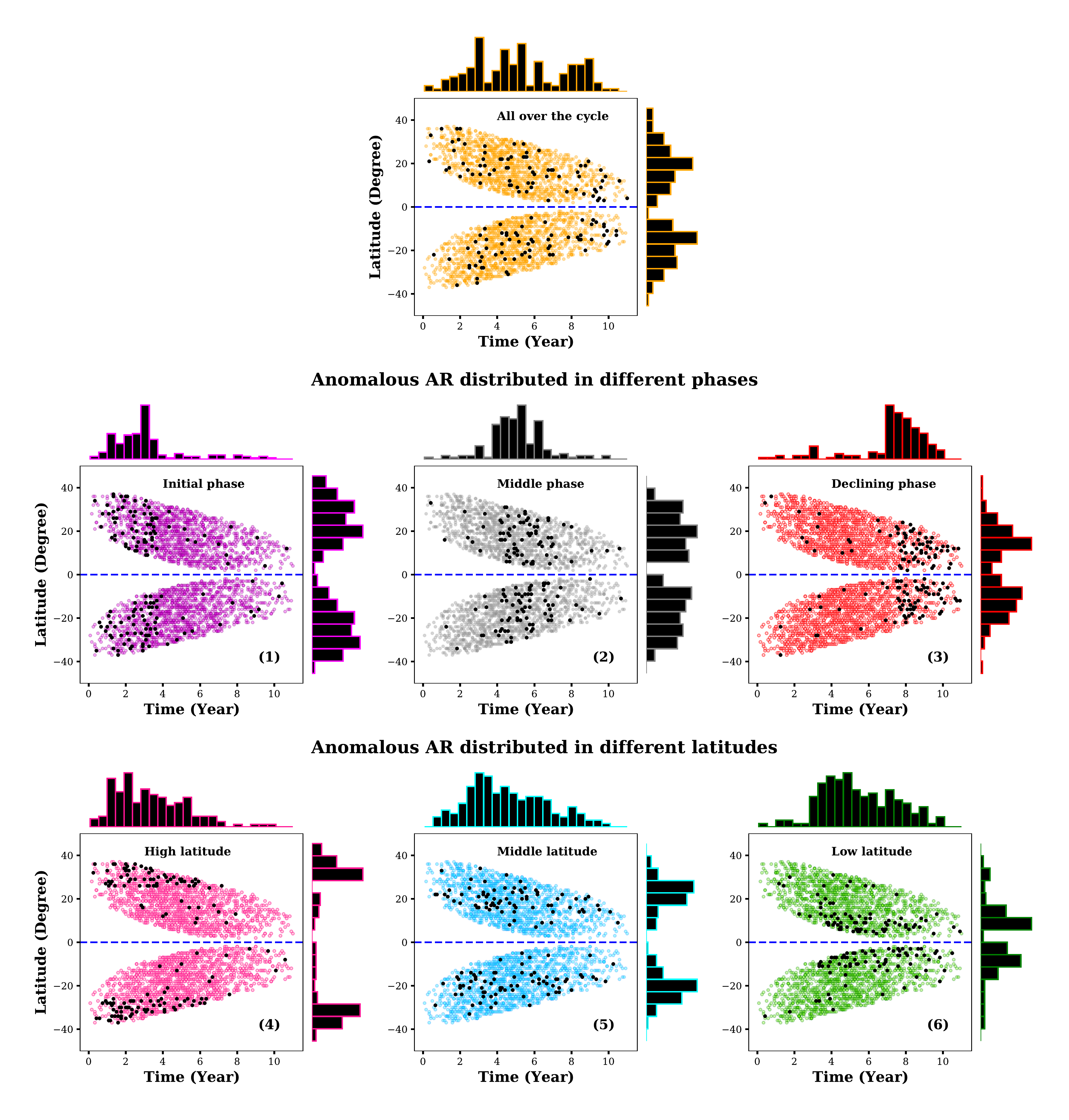}}
\caption{Butterfly diagrams reflecting the spatio-temporal variability of the distribution of anomalous regions. Histograms of phase distributions and latitude distributions are plotted on the top axis and right axis respectively in each subplot. Top panel reflects the distribution where 5$\%$ anomalous regions are spread all over the cycle. The next panel constitutes the diversity in emergence phase of the anomalous active regions -- (1), (2) and (3) dictate population dispersed at initial phase, middle phase and the declining phase, respectively. Third panel denotes the high, mid and low latitude distributions (4, 5 and 6, respectively).}\label{fig:2}
\end{figure}

In most of our simulations, we consider 5$\%$ anomalous regions of any kind (AH-J, H-AJ or AH-AJ) comprising 5$\%$ of the total unsigned flux associated with the active regions of the whole cycle. We place anomalous regions over the time-latitude domain using a random number generator (see the 1st row in Figure~\ref{fig:2}; the black dots represent anomalous regions). In addition, we also prepare six separate cases by varying emergence latitudes and phases associated with AH-J BMRs. The butterfly diagrams corresponding to these cases are shown in the middle and last panel of Figure~\ref{fig:2}. With each of these different synthetic sunspot series, we perform disparate SFT simulations as described in the following section \ref{sec:results}.

\section{Results} \label{sec:results}
\label{sec3}
In order to understand the effect of anomalous active regions on the surface magnetic field we primarily focus on large-scale magnetic field distributions. Especially, it is well-established that the polar field and the axial dipole moment at the end of the solar cycle, which closely reflect the large-scale field distribution, strongly correlate with the succeeding cycle amplitude \citep{Yeates2008,Andres2013,Bhowmik2018}. We calculate the dipole moment in the following way,
\begin{equation}
\mathrm{DM}(t) = \frac{3}{4\pi R_\odot^{2}}\:\int_{\phi=0}^{2\pi}\:\int_{\lambda}^{} \: \:B_r(\lambda,\phi,t)\:\sin\lambda\:\cos\lambda\:d\lambda\:d\phi,
\end{equation}\label{eq5}

\noindent where $\lambda$ and $\phi$ represent latitude and longitude, respectively.  $R_\odot$ is the solar radius. DM(t) denotes the global axial dipole moment. The dipole moment calculation involves the whole latitudinal and longitudinal domain \textit{i.e.} the entire photospheric radial magnetic field.

\subsection{The Effect of Anomalous Regions on Axial Dipole Moment}
\label{sec3.1}

\begin{figure}[h] 
 \centerline{\includegraphics[width=0.8\textwidth,clip=]{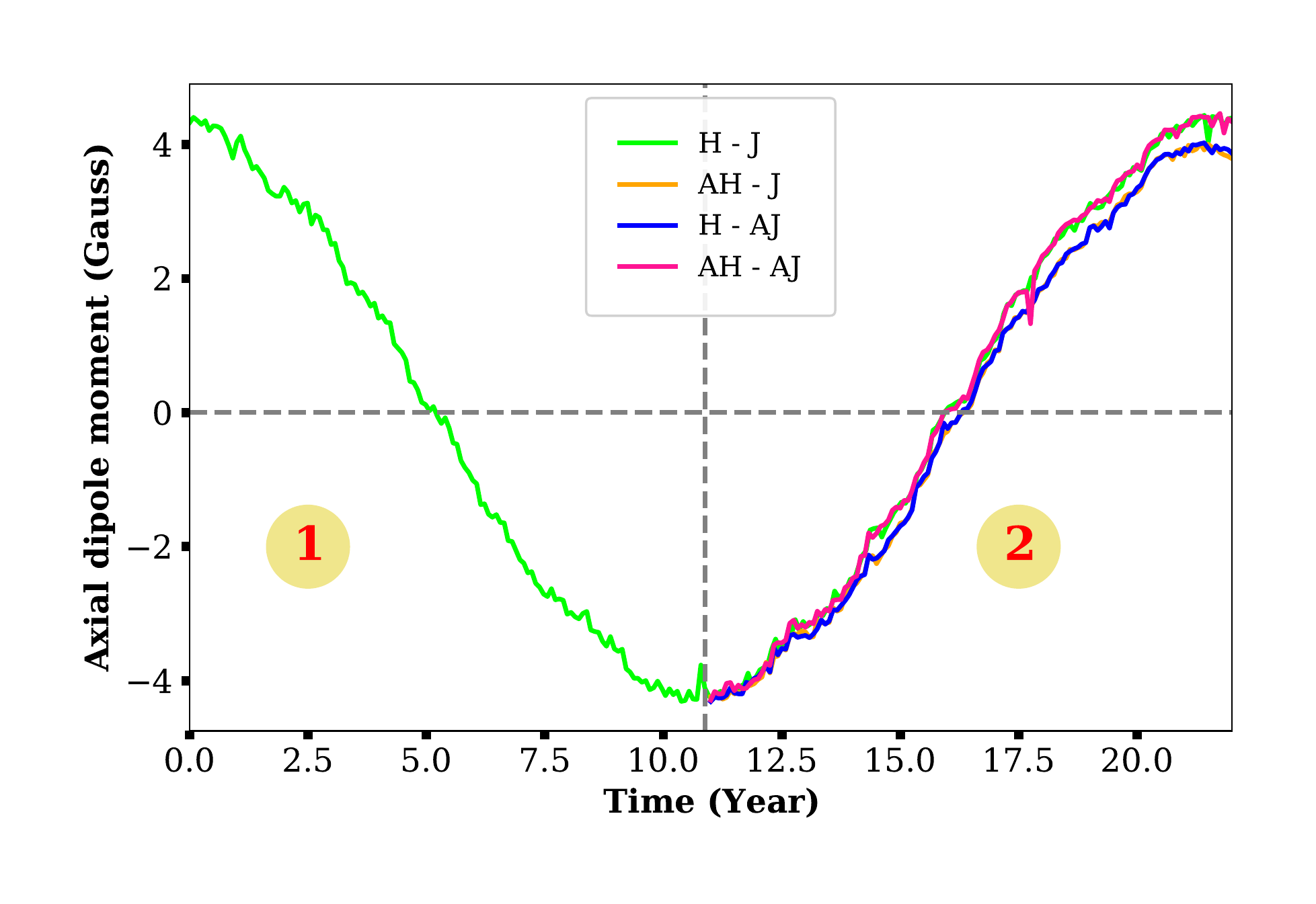}}
\caption{Time variation of axial dipole moment over two solar cycles including 5 $\%$ of four differently configured BMRs. Green and pink curves denote the axial dipole moment evolution of configuration 1 and configuration 4 (AH-AJ), respectively. Orange and blue curves represent the time evolution of dipole moment for configuration 2 (AH-J) and 3 (H-AJ) respectively.}\label{fig:3}
\end{figure}

We initially perform four SFT simulations (duration: 22 years each). One simulation with all H-J spots and the other three simulations are with 5\,$\%$ anomalous regions in configurations AH-J, H-AJ, and AH-AJ respectively, while the remaining 95\,$\%$ are H-J regions. It is noted that the spatio-temporal distribution of anomalous regions is kept the same as sunspots in these cases. Figure~\ref{fig:3} depicts the evolution of the dipole moments associated with each of these four simulations, where the anomalous regions were introduced beyond year 11, \textit{i.e.} in the second cycle. We notice that there is no significant difference in the dipole moment variation for H-J and AH-AJ BMR distribution while AH-J and H-AJ both behave similar to each other, yet significantly different from H-J and AH-AJ. Quantitative details of the dipole moment amplitude for these four configurations are provided in table~\ref{table:1} (see Sl No. 1 to 4).

\begin{figure}[h] 
 \centerline{\includegraphics[width=0.9\textwidth,clip=]{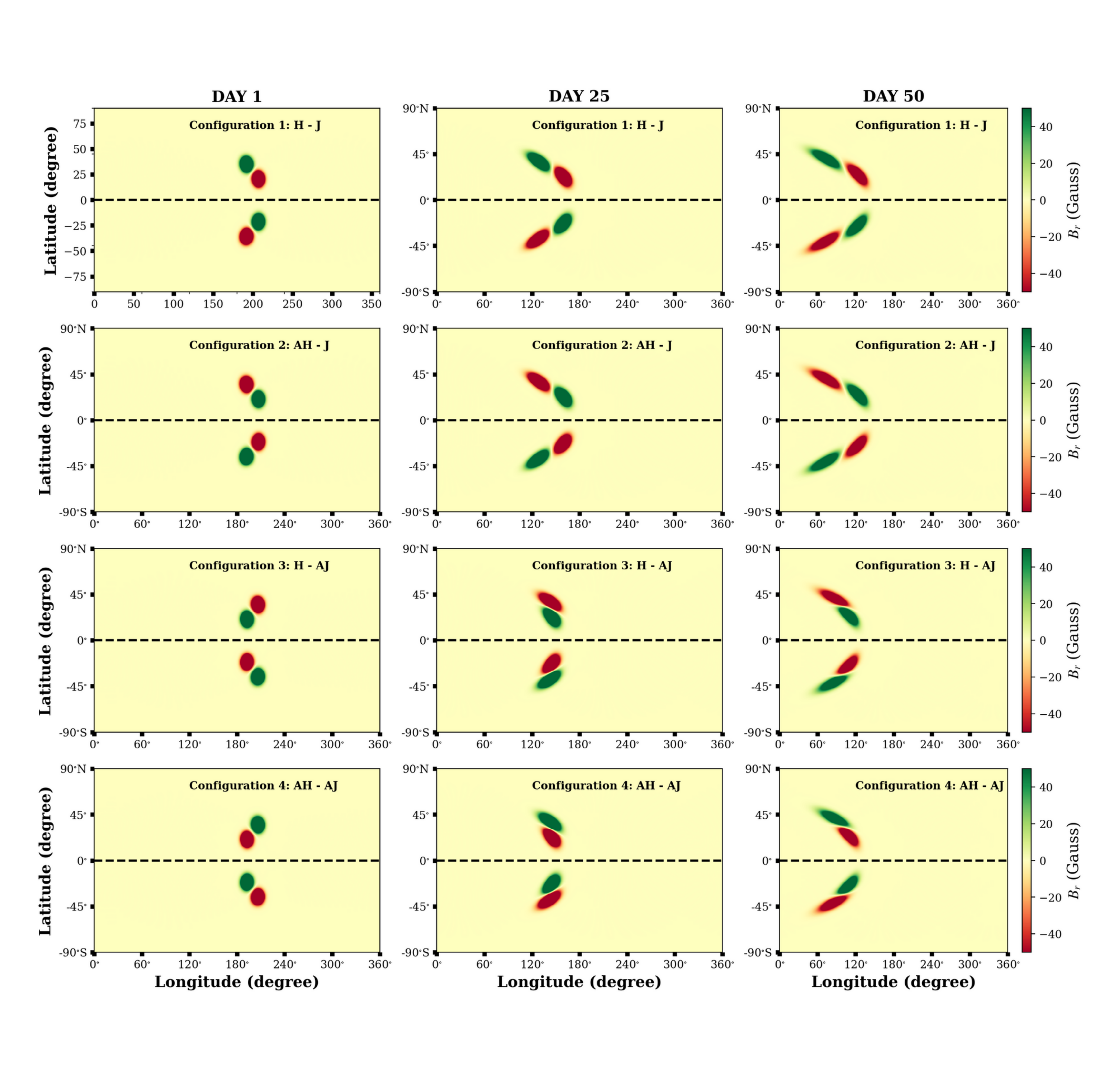}}
\caption{Time evolution of the surface magnetic field ($B_r$) for different configurations. Here configuration 1 stands for H-J bipolar magnetic region (BMR), and configurations 2, 3, and 4 indicates the anomalous regions (mentioned in Figure~\ref{fig:1}). SFT simulations are performed for these four initial configurations, and the evolved magnetic field configuration on day 1, 25 and 50 are shown in the left, middle and right panels, respectively.}\label{fig:4}
\end{figure}

In order to understand what leads to these similar outcomes from apparently different BMR configurations, we investigate the temporal evolution of a single BMR each having $\sim$ ${60}^\circ$ tilt angle corresponding to the four configurations individually (see Figure~\ref{fig:4}). 

Due to differential rotation, the large-scale surface plasma flow in the azimuthal direction is faster near the equator than at higher latitudes. Consequently, the following spot of an H-AJ region nearer to the equator will advect more towards the right than the leading spot and eventually will attain a configuration similar to that of AH-J. After a short period of evolution ($\sim$ 50 days in our simulations), the only difference remaining between an H-AJ and an AH-J region is that H-AJ region has a higher tilt angle and lower separation between its two polarities than AH-J (see the middle two rows in Figure~\ref{fig:4}). The differential rotation has a similar effect on an AH-AJ region, as it evolves to converge toward the standard H-J configuration with a lower latitudinal separation and a higher tilt angle (see the top and bottom rows in Figure~\ref{fig:4}). Now, the timescale of convergence between the two sets of configurations (H-J \& AH-AJ, AH-J \& H-AJ) depends on the initial tilt and emergence latitude, which primarily decide the angular separation between the two spots of the BMR (given that their sizes are similar). In the case of a typical observed active region, its tilt and, thus, the latitudinal separation are much smaller. This makes the timescale of convergence for the two sets of configurations longer. However, the time scale of empirical sunspot evolution is short compared to long-term behaviour and can be considered transient to the entire solar cycle progression. This explains the similar temporal evolution and the final amplitude of the dipole moments at the end of the 22-year-long SFT simulations (H-J -- AH-AJ and AH-J -- AH-AJ), even with apparently different classes of anomalous regions. Increasing the percentage of anomalous regions to as high as 25\,$\%$ does not alter this generic behavior, indicating that the effect of the differential rotation on active regions is quite robust and swift.

\subsection{Effect on the Succeeding Cycle Amplitude and Time Period}
\label{sec3.2}
The previous section demonstrates how the AH-J and H-AJ BMRs effectively cause the same variation in the large-scale dipole moment build-up dynamics. Hereafter we limit our analyses only to AH-J regions for simplicity. In the context of solar cycle predictions using either dipole moment or polar flux, we quantitatively evaluate the effect of AH-J regions on the final dipole moment amplitude at the cycle minimum and speculate its implications on the following cycle amplitude.

\begin{figure}[h] 
 \centerline{\includegraphics[width=1\textwidth,clip=]{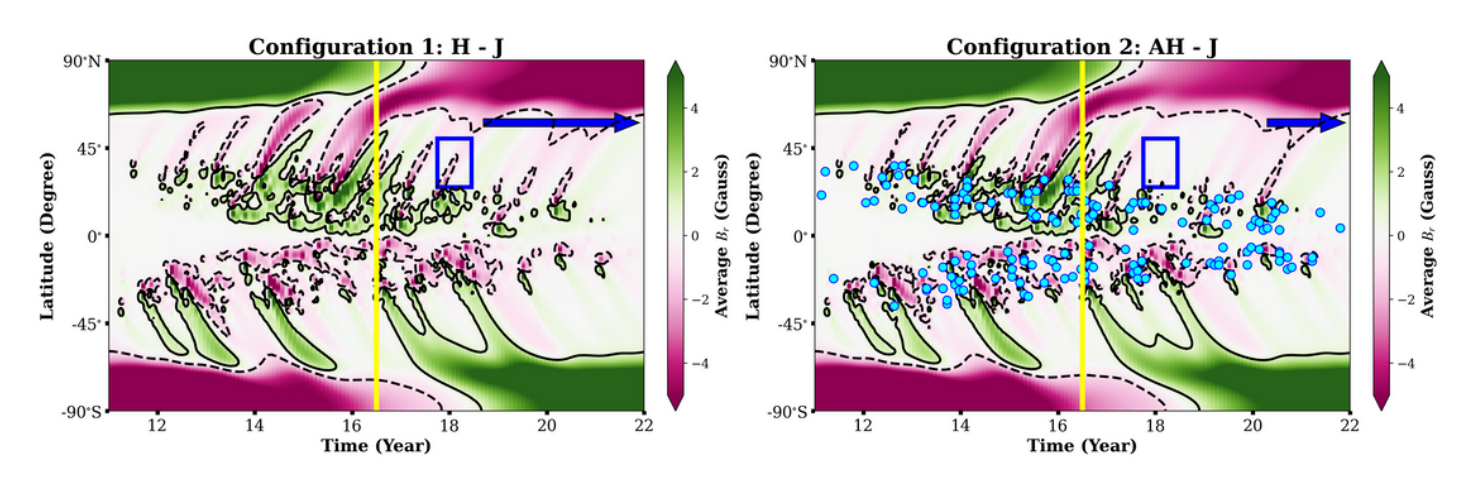}}
\caption{Butterfly diagram of the surface magnetic field ($B_r$) for configuration 1 and configuration 2. \textit{Panel (a):} represents the butterfly diagram of H-J regions. \textit{Panel (b):} shows the butterfly diagram of 95 $\%$ H-J regions with 5 $\%$ AH-J regions. Cyan circles represents the AH-J regions distributed all over the solar cycle. Yellow line in both plots depict the reversal timing of the polar field. Black solid and black dashed lines represent positive and negative polarities respectively. Blue arrow in \textit{Panel (a)} denotes the newly built up polar field region, whereas blue box indicates the opposite polarity surges for H-J BMRs. On the other hand shorter blue arrow in \textit{Panel (b):} reflects the accumulation of lesser amount of negative flux during the end phase of the cycle. Blue box indicates the missing opposite polarity surges in the same panel. }\label{fig:5}
\end{figure}

We utilize magnetic butterfly diagrams (Figure~\ref{fig:5}) depicting the spatio-temporal evolution of the longitudinally averaged $B_r$ on the solar surface to compare two SFT simulations: one with all H-J regions and another one with 5\,$\%$ AH-J regions (remaining 95\,$\%$ are H-J regions). We notice two significant differences.

Firstly, in both hemispheres, magnetic fluxes predominantly from the following spots migrate towards the poles as an effect of the B-L mechanism. The amount of these unipolar surges, negative in the northern hemisphere (positive in the southern hemisphere), is relatively higher in H-J configuration than in AH-J.

Secondly, in the second SFT simulation that contains AH-J regions, some negative (in the northern hemisphere) polar surges are suppressed by the magnetic flux associated with the positive following polarities of AH-J regions (see within the blue rectangle in Figure~\ref{fig:5}). A similar scenario is observed in the southern hemisphere with opposite polarity surges. The overall effect on the evolution and build-up of the large-scale polar magnetic field is manifested in two ways: delay in its polarity reversal and weaker final amplitude at the end of the cycle in the second simulation with AH-J regions compared to purely H-J simulation. Consequently, the succeeding solar cycle is expected to be weaker and have a relatively extended cycle duration \citep{Waldmeier1935,Dikpati2008,Karak2011}. 

Such changes should be more profound with more AH-J spots, which we verify by performing several SFT simulations with increasing percentages of AH-J regions. Statistical correlation analysis of the decrement of dipole moment amplitude at the solar minimum with the increasing percentage of AH-J regions shows a strong positive correlation with 99.9\,$\%$ confidence (Spearman coefficient is 0.963) in our SFT simulations. Subsequently, the delay in dipole moment reversal compared to H-J regions bears a positive correlation with the increasing number of AH-J regions (Spearman coefficient is 0.97 with 99.9\,$\%$ confidence).

\subsection{Relative Effectivity between the Number and Magnetic Flux of Anomalous Regions}
\label{sec3.3}

\begin{figure}[h] 
 \centerline{\includegraphics[width=0.8\textwidth,clip=]{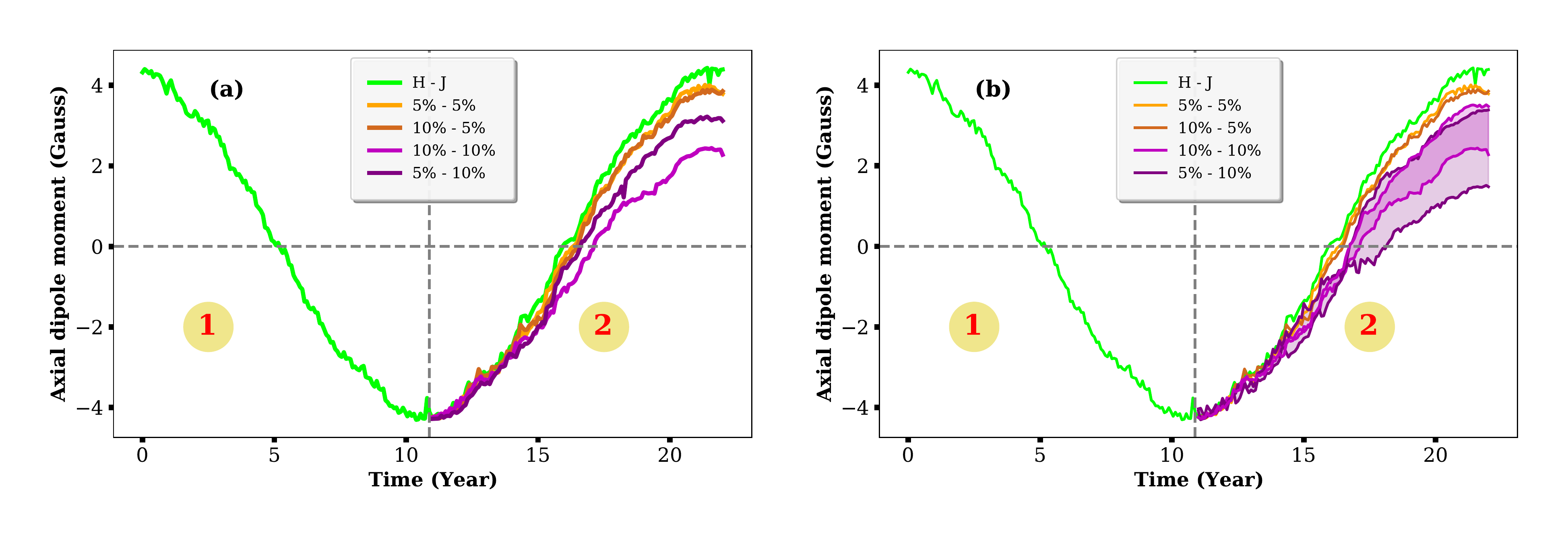}}
\caption{Time evolution of axial dipole moment over 22 years incorporating a variation in the number and flux content of the AH-J regions only in the second cycle. Here the variation in number and flux is mentioned as \%number-\%flux. Orange (5 $\%$-5 $\%$) and dark-orange (10 $\%$-5 $\%$) curves depict the time evolution of dipole moment with AH-J regions having 5 $\%$ flux but 5 $\%$ and 10 $\%$ in number respectively. Violet (10 $\%$-10 $\%$) and dark-violet (5 $\%$-10 $\%$) curves represent AH-J regions having 10$\%$ flux but 5 $\%$ and 10 $\%$ in number respectively. Green curve denotes the corresponding result for $100\%$ H-J regions. \textit{Panel (a):} represents the simulation with single realization. \textit{Panel (b):} depicts the simulations with 20 random realizations for 10 $\%$-10 $\%$ and 5 $\%$-10 $\%$ cases.}\label{fig:6}
\end{figure}

According to the basic features of the B-L mechanism, which dictates the evolution of the large-scale magnetic field on the solar surface, the total amount of magnetic flux associated with active regions should have more significance than the total number of regions itself. Although past observational studies \citep{McClintock2014,Li2018,Andres2021} provide an estimate of the number of AH-J regions (3\,$\%$ to 10\,$\%$), there is scanty information about the total flux contained by these anomalous regions. Thus we perform additional SFT simulations with the aim of determining which one is more important: the number of AH-J regions (relative to H-J regions) or the amount of AH-J region-associated magnetic flux (relative to H-J flux). Thus we consider four distributions and compare the corresponding dipole moment evolution (Figure~\ref{fig:6}). 

\begin{enumerate}[label=(\alph*)]
\item AH-J regions constituting 5\,$\%$ of the total number of active regions and 5\,$\%$ of the total active region-associated magnetic flux. Here the variation in number and flux is mentioned as $\%$\,number-$\%$\,flux (thus, 5\,$\%$\,-\,5\,$\%$ hereafter).
\item AH-J regions constituting 10\,$\%$ of the total active regions and 5\,$\%$ of the total flux (10\,$\%$\,-\,5\,$\%$ hereafter).
\item AH-J regions constituting 5\,$\%$ of the total active regions but 10\,$\%$ of the total flux (5\,$\%$\,-\,10\,$\%$ hereafter).
\item AH-J regions constituting 10\,$\%$ of the total active regions and 10\,$\%$ of the total flux (10\,$\%$\,-\,10\,$\%$ hereafter).
\end{enumerate}

In Figure~\ref{fig:6}a, we first observe a close resemblance between the dipole moment evolution in cases with fixed flux but different numbers of AH-J regions: 5\,$\%$\,-\,5\,$\%$ (orange curve) and 10\,$\%$\,-\,5\,$\%$ (dark-orange curve). Whereas changing the flux from 5\,$\%$ to 10\,$\%$ while keeping the number of AH-J regions fixed to 5\,$\%$ shows a more profound impact on the reversal timing and final amplitude of the dipole moment at the end of the cycle (see the orange and dark-orange curve in Figure~\ref{fig:6}a). This difference in the large-scale magnetic field evolution becomes increasingly apparent in the later half of the cycle. However, the importance of the number density of AJ-H regions cannot be entirely disregarded. The same amount of magnetic flux (for example, 10\,$\%$) can be distributed in one or many AH-J regions randomly positioned in the activity belt. Depending on their emergence timing and latitude (and tilt angle), additional variations will appear in the dipole moment evolution. We test it by performing additional twenty SFT simulations with different realizations of the time-latitudinal distribution of AH-J regions. The corresponding dipole moment evolution is depicted in Figure~\ref{fig:6}b. The outer curves encompassing the shaded area denote the dipole moment evolution with maximum and minimum amplitudes for cases: 5\,$\%$\,-\,10\,$\%$ (dark-magenta) and 10\,$\%$\,-\,10\,$\%$ (magenta). The width and significant overlap of these two regions indicate that the relative importance between the number and magnetic flux is subjected to the particular distribution of the associated emergence timing, location and latitude-based tilt angles of anomalous spots. The details of these variations are discussed in the next section.

\subsection{Effect of Spatio-temporal Variability of the Anomalous Active Region Distribution}
\label{sec3.4}

So far, in our SFT simulations, we randomly distribute the anomalous regions over the activity belt of the entire solar cycle. However, results from the previous section (\ref{sec3.3}) indicate that emergence timing and latitude of such regions could introduce more variability in the large-scale magnetic field evolution. Moreover, earlier studies on observed sunspots \citep{Sokoloff2015,Li2018} claim preferential spatio-temporal distribution of anomalous regions which differs from cycle to cycle. Thus, we perform six more SFT simulations to address variations of AH-J region distribution as follows.

\begin{figure}[h] 
 \centerline{\includegraphics[width=0.8\textwidth,clip=]{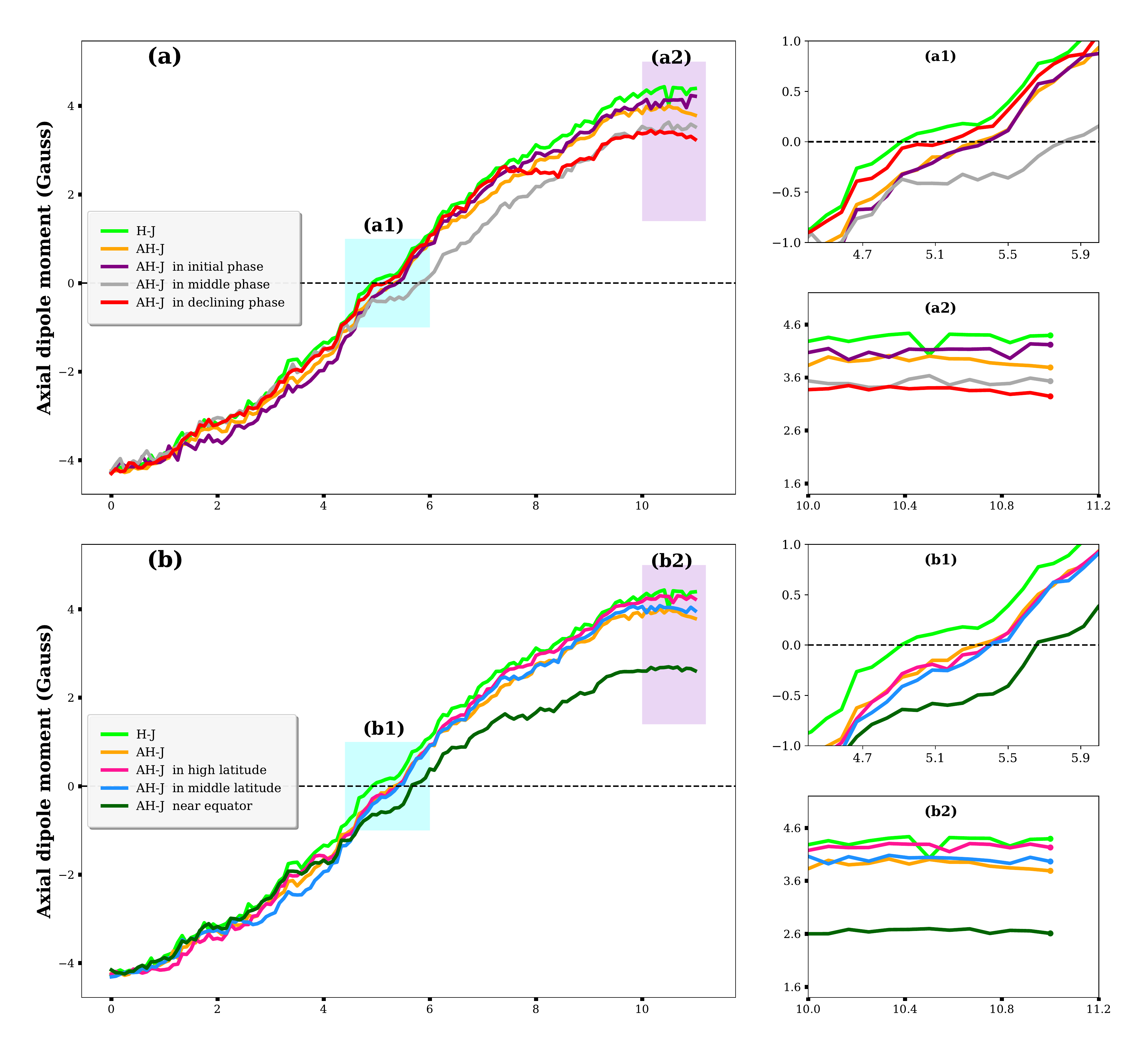}}
\caption{Variation in the time series of the dipole moment (only for the second cycle in the simulation) with different distribution of anomalous (AH-J) regions shown in \ref{fig:2}. \textit{Panel (a):} represents the dipole moment evolution of different phase distributions. The effect of AH-J regions having more density near the starting of the cycle, middle of the cycle and the end of the cycle on the dipole moment is denoted by violet, grey and red curves, respectively. \textit{Panel (b):} pink, light blue and dark green curves depict the dipole moment evolution for AH-J regions emerging in high, mid and low latitudinal positions, respectively.
In all of these plots, green and orange curves represent H-J and  AH-J, respectively. Panel a1, b1 and a2, b2 are the cut out regions of panels a, b highlighted in cyan and violet, respectively. These sub-panels portray the delay in time-reversal near the cycle maximum and the decrement of dipole moment/polar flux at the minimum of the same cycle.}\label{fig:7}
\end{figure}

We consider that AH-J regions emerging during the: (1) initial phase of the cycle, (2) middle phase i.e. around the cycle maximum, and (3) declining phase of the cycle. Furthermore, we consider other cases with AH-J regions concentrated (4) at higher latitudes, (5) at mid-latitudes and (6) near the equatorial region (low latitudes). The corresponding activity belts are shown in Figure~\ref{fig:2} and the dipole moment evolution is depicted in Figure~\ref{fig:7}.

For the case with AH-J regions emerging predominantly in the initial phase (simulation 1) of the cycle, we notice a slightly delayed reversal (Figure~\ref{fig:7}a) of the dipole moment caused by unipolar surges from the following spots of AH-J regions which have the same polarity as the existing polar field. However, in the later stage of the cycle, the surface field is dictated by mostly H-J regions, which facilitate a faster increase of the dipole moment, thereby resulting in a final amplitude similar to the standard randomly distributed AH-J regions all over the activity belt (see orange and violet curve in Figure~\ref{fig:7}a). The impact of AH-J regions appearing around the cycle maximum (simulation 2) is reflected in the significant delay in the reversal of dipole moment (see the grey curve in Figure~\ref{fig:7}a). Note that the bulk of AH-J regions are introduced from year 4 onwards, and the dipole moment almost immediately starts deviating from the H-J case showing a rather sharp deviation in year 5. The effect persists until the cycle minimum with a weaker final dipole moment. Finally, in the case of AH-J regions emerging during the declining phase of the cycle (simulation 3), the dipole moment reversal epoch is similar to the H-J simulation (see the green and red curves in Figure~\ref{fig:7}a). However, the final dipole moment strength is the weakest among all three cases mentioned so far. It is caused by the magnetic flux contribution from the following spots of AH-J regions (having opposite polarity compared to the newly-built polar field) appearing mostly at the end of the activity cycle.

The dipole moment evolution for the three simulations with AH-J regions preferentially emerging in the high (simulation 4), mid (simulation 5), and low latitudes (simulation 6) are depicted in Figure~\ref{fig:7}b. We notice that the high-latitude distribution has minimal effect on the dipole moment (pink curve in Figure~\ref{fig:7}b), resulting in a similar evolution as in all H-J regions case, also concurring with the findings of \cite{Yeates2015}. However, the dipole moment corresponding to the simulation with mid-latitude emergences of AH-J regions is almost equivalent to the standard case (where AH-J regions are distributed all over the cycle). The maximum impact on the dipole moment is caused by the AH-J regions emerging at low latitudes (dark green curve in Figure~\ref{fig:7}b), which is directly related to their negative contribution (higher than the other cases) to the dipole moment build-up (explained in more detail in the following sections). Quantitative details on how the dipole moment evolution is affected in these six cases are provided in table~\ref{table:1} (see Sl No. 5 to 10).

\section{Effectivity of Anomalous Regions on Global Dipole Moment: A Theoretical Perspective}
\label{theory}
In our analyses so far, we have considered the large-scale magnetic field evolution on the solar surface, primarily the dipole moment and polar flux, both of which are longitudinally invariant by definition. While focusing only on these two measures, our spatially two-dimensional SFT simulation (function of latitude, longitude and time) can be reduced to only one spatial dimension with surface magnetic field dependent on latitude and time \citep{Petrovay2020}. Under such simplification, each tilted BMR, after averaging azimuthally, will appear as a pair of flux rings of opposite polarity with a bipole source having a finite latitudinal separation. This methodology has been utilized earlier \citep{Petrovay2020,Nagy2020,Yeates2020} to propose a mathematical perspective of calculating the individual contribution of each active region emerging during a solar cycle to the final global dipole moment generation at the cycle minimum.

As demonstrated by \cite{Petrovay2020} and \cite{Nagy2020}, the change in the global dipole moment during the n$^{\mathrm{th}}$ cycle can be expressed as the sum of the contributions from individual active region dipole moments as follows,

\begin{equation}
    \mathrm{DM}_{n+1} - \mathrm{DM}_n = \Delta \mathrm{DM} =  \sum_{i=1}^{N_{\mathrm{total}}} \delta D_{\mathrm{U},i}  
\end{equation}

\noindent where $\delta D_{\mathrm{U},i}$ corresponds to the `ultimate' dipole moment contribution from the \textit{i}$^{\mathrm{th}}$ active region during n$^{\mathrm{th}}$ cycle and assuming there are $N_\mathrm{total}$ regions in that cycle. Now, the final contribution, $\delta D_{\mathrm{U},i} = f_{\infty,i}\,\delta D_{1,i}$, where $f_{\infty, i}$ is the asymptotic factor \citep{Petrovay2020}. The initial unsigned dipole moment of any active region can be expressed as $\delta D_1 = \frac{3}{4 \pi R^2}\,\Phi\,d_\lambda \cos \lambda$, where $\lambda$ is the emergence latitude and $d_\lambda$ is the latitudinal separation of the two polarities of the sunspot. $\Phi$ represents the magnetic flux in the BMR's leading (or following) spot. The polarity of the spots closest to the equator determines whether $\delta D_1$ will positively or negatively contribute to the global dipole moment. 

The asymptotic factor is a Gaussian function of latitude, $f_{\infty} = \mathrm{C}\,\mathrm{exp}(- \lambda^2/ 2\,\lambda_R^2)$, such that its amplitude decreases with increasing latitude. $\lambda_R$ and C are dependent only on the transport parameters used in the SFT model, which are fixed in all our simulations. Thus, in summary, the latitude dependency of the ultimate dipole moment contribution from an individual region can be expressed as,

\begin{equation}
    \delta D_\mathrm{U} \propto d_{\lambda}\, \cos \lambda \,\mathrm{e}^{-\lambda^2}.  
    \label{eq11}
\end{equation}

\begin{figure}[h] 
 \centerline{\includegraphics[width=0.9\textwidth,clip=]{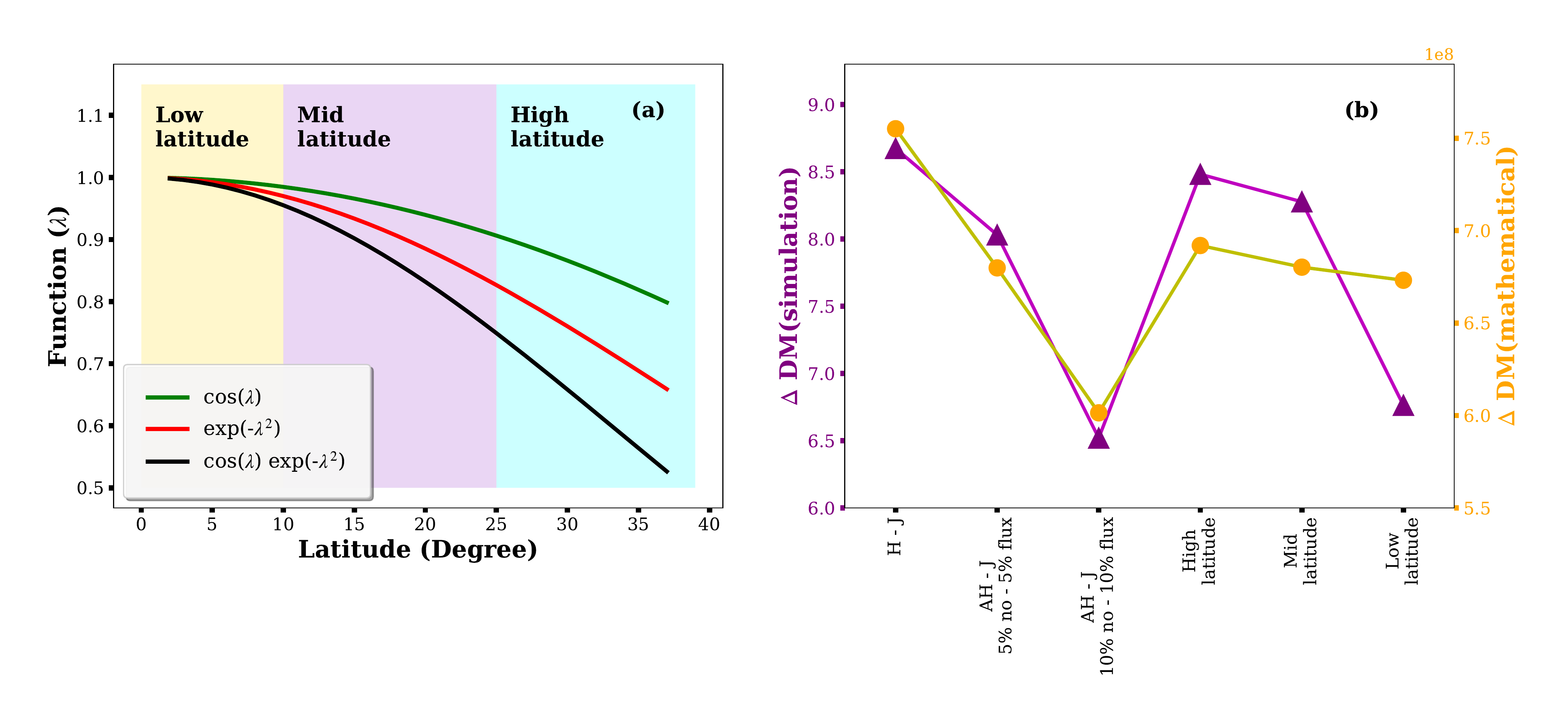}}
\caption{Qualitative comparison of net change in solar dipole moment in our SFT simulation with the mathematical approximations. \textit{Panel (a):} depicts the latitudinal ($\lambda$) dependency of different analytical functions contributing to the ultimate dipole moment. High, mid and low latitudinal regions are denoted by cyan, violet and yellow. \textit{Panel (b):} change in dipole moment ($\Delta$ DM) of our simulations(violet traingles) are compared with the mathematically derived approximated values (orange circles) for six different cases labelled on the horizontal axis. }\label{fig:8}
\end{figure}

The function, $\delta D_{\mathrm{U}}$ has decaying amplitude with increasing latitude, provided $d_\lambda$ is constant for a 1D system (see Figure~\ref{fig:8}a). This theoretical aspect delineated above can be utilized to explain the variations seen in the dipole moment evolution in multiple SFT simulations presented in section \ref{sec3}. Firstly, under this one-dimensional formulation, active regions (in Figure~\ref{fig:2}) are reduced to the bipole sources with finite latitudinal separation, whereas the longitudinal separation between the leading and following spots becomes irrelevant. Thus we find pair-wise similarities of dipole moment evolution for the cases: H-J and AH-AJ, and AH-J and H-AJ (provided all emerge at the same latitude, $\lambda$ with the same separation, $d_\lambda$). Secondly, any AH-J (or H-AJ) region will contribute negatively to the dipole moment as its relative polarity configuration is the exact opposite of an H-J (or AH-AJ) region, thus diminishing the global dipole moment as found in sections \ref{sec3.1} and \ref{sec3.2}. These findings are reflected in Figure~\ref{fig:8}b where we carry out a qualitative comparison between the net change in solar dipole moment in SFT simulations with the mathematical approximations.

The latitude dependency shown in equation \ref{eq11} also supports our findings in section \ref{sec3.4}. In the first three SFT simulations with the varying spatio-temporal distribution of AH-J regions on the activity belts, we see notable changes in global dipole moment in cases with the anomalous regions emerging during either middle or declining phases of the cycle (see Figure~\ref{fig:7}a). A careful inspection of the latitudinal distribution of these spots on the activity belt (see the second row in Figure~\ref{fig:4}) suggests that more spots emerging in mid-to-low latitudes (in contrast to more high latitude emergences) are responsible for this. Similarly, for cases with AH-J regions appearing in different latitudinal belts (see the third row in Figure~\ref{fig:4}), preferential emergence in low latitudes (smaller $\lambda$) will increase the negative contribution according to equation \ref{eq11}. Therefore, it will cause the maximum reduction of the final global dipole moment (see the pink curve in Figure~\ref{fig:7}b). Figure~\ref{fig:8}b shows the qualitative comparison between these results (high/mid/low latitudinal emergence impacts) from our SFT simulations and aforementioned algebraic prescription.

\section{Concluding Discussions} \label{sec:conclusion}

Emergence of active regions and transport of the associated magnetic flux on the solar surface primarily determines the Sun's polar field and dipole moment build-up. The global dipolar field acts as the seed which modulate the amplitude of the future sunspot cycle underscoring its importance. In this study, we focus on the orientation of the active regions and investigate their impacts on the large-scale magnetic field of the Sun, especially, the dipole moment and hemispheric polar field. We keep the flux transport parameters and some BMR properties like the separation, area, latitudinal and longitudinal position fixed, eliminating variations originating from these properties. This enables us to segregate contributions from ``anomalous sunspot regions" in the creation of irregularities in the solar cycle. It is to be noted that ``rogue'' regions \citep{Nagy2017} with very high tilt angle or high magnetic flux (or a combination of both) are a subset of anomalous sunspots according to our classification. How a few rogue or highly tilted active regions can significantly impact the dipole moment evolution has been studied earlier \citep{jiang2014b, Nagy2017}. In comparison, our focus is not on rogue regions specifically but on a more comprehensive study of the impact of diverse classes of anomalous regions with varying numbers, flux content emergence timing and location.

We consider all active regions as bipolar sunspots (i.e., $\beta$-spots) and do not consider any regions with complex morphology (e.g., $\delta$-spots). By performing multiple SFT simulations we find that a fraction of the total sunspot number emerging as anomalous regions can significantly influence the large-scale magnetic field evolution during the solar cycle. The appearance of AH-J/H-AJ regions changes the polarity reversal timing and eventually suppresses the axial dipole moment at the end of the solar cycle -- in conformity with the findings of \citep{Jiang2015,Nagy2017}. 

In this work, we establish a crucial point through SFT simulations: the AH-J and H-AJ regions contribute almost identically in the dipole moment evolution, polar field build-up and their reversal epoch; this is because over a few rotational timescales the Sun's differential rotation makes their orientation similar. In addition, AH-AJ active regions behave similarly to H-J regions in their long-term evolution. We perform an analysis which shows that our simulation results are consistent with the algebraic formulation motivated from mathematical theory (see section \ref{theory}).
 
Based on our findings we conclude that in the context of the the long-term evolution of the Sun's large-scale magnetic field, the sunspots' orientation can be grouped into two classes: 1) Those that contribute positively to the dipole moment -- Hale-Joy sunspots (H-J) and 2) Which contribute negatively -- Anti Hale-Joy sunspots (AH-J). These two distinct classes of active region configurations also exhibit characteristically distinct dynamics over shorter time scales leading to disparate interactions with the local magnetic field distribution. 

Our analyses demonstrate that the amount of flux and the number density of anomalous regions play an essential role in the dynamics of surface magnetic field evolution. Although not always, we notice that in some cases a large population of anomalous sunspots can impact the ultimate dipole moment more than a small population carrying the same amount of flux. Therefore we speculate that sometimes a larger group of anomalous spots can cumulatively be more influential than a single large rogue region due to their latitudinal and phase distribution diversity.

It is known that the total amount of unsigned polar flux is of the order of the unsigned flux contained in a single but very large active region. A similar amount of net flux distributed over a fraction of sunspots of the anomalous class of regions appearing at distinct phases and latitudes will affect the cycle differently. Our simulations show that AH-J regions near the equator have the highest impact on the ultimate dipole moment strength, whereas their appearance during the mid-phase of the solar cycle has the highest impact on the reversal timing. Near-equatorial emergence is understood to be more influential in general. This carries over to the case of anomalous regions. The leading polarity of AH-J/H-AJ regions is opposite to the sign of the new polar field (that is being imparted) in the opposite hemisphere. Thus upon appear near the equator and diffusing to the other hemisphere, it may reduce the strength of the new polar field of the opposite hemisphere (and the ultimate dipole moment) during the descending phase of the solar cycle. In sharp contrast, at high latitudes, the initial phase population of anomalous regions hardly influences the overall magnetic cycle evolution and its inherent dynamics. We believe that along with the diverse phase and latitudinal distributions, the hemispheric asymmetry in the distribution of AH-J/H-AJ regions may translate to the asymmetry seen in the polar field weakening.

The result from different scenarios based on different synthetic sunspot distributions is described in section \ref{sec:results} and tabulated in table \ref{table:1}. Note that the statistics reported correspond to one random realization of each scenario. Although these values may differ from one random realization to the other, the qualitative behaviour of the global magnetic field evolution under different possible scenarios still have a resemblance to the observed Sun.

In summary, our simulations provide insight on how the interplay of anomalous and regular active regions modulate the solar dipole moment evolution. Given that the dipole moment at the minima of a cycle is the most important factor determining the strength of the future sunspot cycle, these insights are important in the context of solar cycle predictions. In order to constrain polar field evolution and dynamics it would be important to extract the combined data of AH-J, H-AJ and AH-AJ regions (with their precise number, flux content and the emergence phase-latitude information) along with the H-J regions from observation. Such an exercise will provide the necessary observational constraints for driving more precise predictive models of surface flux transport evolution and illuminate the fine subtleties of the Babcock-Leighton mechanism for solar polar field production.

\section{Acknowledgement}
The Center of Excellence in Space Sciences India (CESSI) -- where this work was carried out -- is funded by IISER Kolkata, Ministry of Education, Government of India. S.P. acknowledges funding from the University Grants Commission, Government of India. P.B. acknowledges support from the project ST/W00108X/1 funded by the Science and Technology Facilities Council (STFC), United Kingdom and S.S.M. acknowledges support from NASA contract NAS5-02139 (HMI) to Stanford University. The authors acknowledge useful discussions at ISSI Team Meetings on ``What Determines The Dynamo Effectivity Of Solar Active Regions?'' supported by the International Space Science Institute, Bern, Switzerland. The authors thank the referee for useful discussions.

\bibliographystyle{aasjournal}
%\nocite{*}
\bibliography{anomaly_reg}  

\begin{thebibliography}{}
\expandafter\ifx\csname natexlab\endcsname\relax\def\natexlab#1{#1}\fi
\providecommand{\url}[1]{\href{#1}{#1}}
\providecommand{\dodoi}[1]{doi:~\href{http://doi.org/#1}{\nolinkurl{#1}}}
\providecommand{\doeprint}[1]{\href{http://ascl.net/#1}{\nolinkurl{http://ascl.net/#1}}}
\providecommand{\doarXiv}[1]{\href{https://arxiv.org/abs/#1}{\nolinkurl{https://arxiv.org/abs/#1}}}

\bibitem[{{Babcock}(1961)}]{Babcock1961}
{Babcock}, H.~W. 1961, \apj, 133, 572, \dodoi{10.1086/147060}

\bibitem[{{Bhowmik}(2019)}]{Bhowmik2019}
{Bhowmik}, P. 2019, \aap, 632, A117, \dodoi{10.1051/0004-6361/201834425}

\bibitem[{{Bhowmik} {et~al.}(2023){Bhowmik}, {Jiang}, {Upton}, {Lemerle}, \&
  {Nandy}}]{Bhowmik2023}
{Bhowmik}, P., {Jiang}, J., {Upton}, L., {Lemerle}, A., \& {Nandy}, D. 2023,
  arXiv e-prints, arXiv:2303.12648, \dodoi{10.48550/arXiv.2303.12648}

\bibitem[{{Bhowmik} \& {Nandy}(2018)}]{Bhowmik2018}
{Bhowmik}, P., \& {Nandy}, D. 2018, Nature Communications, 9, 5209,
  \dodoi{10.1038/s41467-018-07690-0}

\bibitem[{{Cameron} {et~al.}(2010){Cameron}, {Jiang}, {Schmitt}, \&
  {Sch{\"u}ssler}}]{Cameron2010}
{Cameron}, R.~H., {Jiang}, J., {Schmitt}, D., \& {Sch{\"u}ssler}, M. 2010,
  \apj, 719, 264, \dodoi{10.1088/0004-637X/719/1/264}

\bibitem[{Carrington(1858)}]{Carrington1858}
Carrington, R.~C. 1858, Monthly Notices of the Royal Astronomical Society, 19,
  1, \dodoi{10.1093/mnras/19.1.1a}

\bibitem[{{Charbonneau}(2020)}]{Charbonneau2020}
{Charbonneau}, P. 2020, Living Reviews in Solar Physics, 17, 4,
  \dodoi{10.1007/s41116-020-00025-6}

\bibitem[{{Choudhuri} \& {Gilman}(1987)}]{Choudhuri1987}
{Choudhuri}, A.~R., \& {Gilman}, P.~A. 1987, \apj, 316, 788,
  \dodoi{10.1086/165243}

\bibitem[{{Dash} {et~al.}(2020){Dash}, {Nandy}, \& {Pal}}]{DASH2020AGU}
{Dash}, S., {Nandy}, D., \& {Pal}, S. 2020, in AGU Fall Meeting Abstracts, Vol.
  2020, SH014--06

\bibitem[{{Dasi-Espuig} {et~al.}(2010){Dasi-Espuig}, {Solanki}, {Krivova},
  {Cameron}, \& {Pe{\~n}uela}}]{Dasi2010}
{Dasi-Espuig}, M., {Solanki}, S.~K., {Krivova}, N.~A., {Cameron}, R., \&
  {Pe{\~n}uela}, T. 2010, \aap, 518, A7, \dodoi{10.1051/0004-6361/201014301}

\bibitem[{Dikpati {et~al.}(2006)Dikpati, de~Toma, \& Gilman}]{Dikpati2006}
Dikpati, M., de~Toma, G., \& Gilman, P.~A. 2006, Geophysical Research Letters,
  33, \dodoi{https://doi.org/10.1029/2005GL025221}

\bibitem[{Dikpati {et~al.}(2008)Dikpati, Gilman, \& de~Toma}]{Dikpati2008}
Dikpati, M., Gilman, P.~A., \& de~Toma, G. 2008, The Astrophysical Journal,
  673, L99, \dodoi{10.1086/527360}

\bibitem[{Durney(1974)}]{Durney1974}
Durney, B. 1974, Solar Physics, 38, \dodoi{10.1007/BF00155068}

\bibitem[{{Fan}(2021)}]{Fan2021}
{Fan}, Y. 2021, Living Reviews in Solar Physics, 18, 5,
  \dodoi{10.1007/s41116-021-00031-2}

\bibitem[{{Fisher} {et~al.}(1995){Fisher}, {Fan}, \& {Howard}}]{Fisher1995}
{Fisher}, G.~H., {Fan}, Y., \& {Howard}, R.~F. 1995, \apj, 438, 463,
  \dodoi{10.1086/175090}

\bibitem[{{Gilman}(2018)}]{Gilman2018}
{Gilman}, P.~A. 2018, \apj, 853, 65, \dodoi{10.3847/1538-4357/aaa4f4}

\bibitem[{{Hale} {et~al.}(1919){Hale}, {Ellerman}, {Nicholson}, \&
  {Joy}}]{Hale1919}
{Hale}, G.~E., {Ellerman}, F., {Nicholson}, S.~B., \& {Joy}, A.~H. 1919, \apj,
  49, 153, \dodoi{10.1086/142452}

\bibitem[{{Hale} \& {Nicholson}(1925)}]{Hale1925}
{Hale}, G.~E., \& {Nicholson}, S.~B. 1925, \apj, 62, 270,
  \dodoi{10.1086/142933}

\bibitem[{{Hanasoge}(2022)}]{Shravan2022}
{Hanasoge}, S.~M. 2022, Living Reviews in Solar Physics, 19, 3,
  \dodoi{10.1007/s41116-022-00034-7}

\bibitem[{{Hathaway}(1993)}]{Hathaway1993}
{Hathaway}, D.~H. 1993, in Astronomical Society of the Pacific Conference
  Series, Vol.~42, GONG 1992. Seismic Investigation of the Sun and Stars, ed.
  T.~M. {Brown}, 265

\bibitem[{{Hathaway}(2010)}]{Hathaway2010}
{Hathaway}, D.~H. 2010, Living Reviews in Solar Physics, 7, 1,
  \dodoi{10.12942/lrsp-2010-1}

\bibitem[{{Hazra} {et~al.}(2023){Hazra}, {Nandy}, {Kitchatinov}, \&
  {Choudhuri}}]{Hazra2023}
{Hazra}, G., {Nandy}, D., {Kitchatinov}, L., \& {Choudhuri}, A.~R. 2023, arXiv
  e-prints, arXiv:2302.09390, \dodoi{10.48550/arXiv.2302.09390}

\bibitem[{Hazra \& Nandy(2016)}]{Hazra2016}
Hazra, S., \& Nandy, D. 2016, The Astrophysical Journal, 832, 9,
  \dodoi{10.3847/0004-637X/832/1/9}

\bibitem[{{Jiang} {et~al.}(2011){Jiang}, {Cameron}, {Schmitt}, \&
  {Sch{\"u}ssler}}]{Jiang2011}
{Jiang}, J., {Cameron}, R.~H., {Schmitt}, D., \& {Sch{\"u}ssler}, M. 2011,
  \aap, 528, A82, \dodoi{10.1051/0004-6361/201016167}

\bibitem[{{Jiang} {et~al.}(2015){Jiang}, {Cameron}, \&
  {Sch{\"u}ssler}}]{Jiang2015}
{Jiang}, J., {Cameron}, R.~H., \& {Sch{\"u}ssler}, M. 2015, \apjl, 808, L28,
  \dodoi{10.1088/2041-8205/808/1/L28}

\bibitem[{Jiang {et~al.}(2014)Jiang, Cameron, \& Schüssler}]{Jiang2014a}
Jiang, J., Cameron, R.~H., \& Schüssler, M. 2014, The Astrophysical Journal,
  791, 5, \dodoi{10.1088/0004-637x/791/1/5}

\bibitem[{{Jiang} {et~al.}(2014){Jiang}, {Hathaway}, {Cameron}, {Solanki},
  {Gizon}, \& {Upton}}]{jiang2014b}
{Jiang}, J., {Hathaway}, D.~H., {Cameron}, R.~H., {et~al.} 2014, \ssr, 186,
  491, \dodoi{10.1007/s11214-014-0083-1}

\bibitem[{Karak \& Choudhuri(2011)}]{Karak2011}
Karak, B.~B., \& Choudhuri, A.~R. 2011, Monthly Notices of the Royal
  Astronomical Society, 410, 1503, \dodoi{10.1111/j.1365-2966.2010.17531.x}

\bibitem[{Kleeorin {et~al.}(2020)Kleeorin, Safiullin, Kuzanyan, Rogachevskii,
  Tlatov, \& Porshnev}]{Kleeorin2020}
Kleeorin, N., Safiullin, N., Kuzanyan, K., {et~al.} 2020, Monthly Notices of
  the Royal Astronomical Society, 495, 238, \dodoi{10.1093/mnras/staa1047}

\bibitem[{{Knizhnik} {et~al.}(2018){Knizhnik}, {Linton}, \&
  {DeVore}}]{Knizhnik2018}
{Knizhnik}, K.~J., {Linton}, M.~G., \& {DeVore}, C.~R. 2018, \apj, 864, 89,
  \dodoi{10.3847/1538-4357/aad68c}

\bibitem[{{Komm} {et~al.}(1993){Komm}, {Howard}, \&
  {Harvey}}]{1993SoPh..147..207K}
{Komm}, R.~W., {Howard}, R.~F., \& {Harvey}, J.~W. 1993, \solphys, 147, 207,
  \dodoi{10.1007/BF00690713}

\bibitem[{{Kumar, Rohit} {et~al.}(2019){Kumar, Rohit}, {Jouve, Laur\`ene}, \&
  {Nandy, Dibyendu}}]{Kumar2019}
{Kumar, Rohit}, {Jouve, Laur\`ene}, \& {Nandy, Dibyendu}. 2019, A\&A, 623, A54,
  \dodoi{10.1051/0004-6361/201834705}

\bibitem[{{Leighton}(1964)}]{Leighton1964}
{Leighton}, R.~B. 1964, \apj, 140, 1547, \dodoi{10.1086/148058}

\bibitem[{Li(2018)}]{Li2018}
Li, J. 2018, \apj, 867, 89, \dodoi{10.3847/1538-4357/aae31a}

\bibitem[{{Mackay} \& {Yeates}(2012)}]{Mackay2012}
{Mackay}, D.~H., \& {Yeates}, A.~R. 2012, Living Reviews in Solar Physics, 9,
  6, \dodoi{10.12942/lrsp-2012-6}

\bibitem[{Mahajan {et~al.}(2021)Mahajan, Hathaway, Mu{\~{n}}oz-Jaramillo, \&
  Martens}]{Mahajan2021}
Mahajan, S.~S., Hathaway, D.~H., Mu{\~{n}}oz-Jaramillo, A., \& Martens, P.~C.
  2021, The Astrophysical Journal, 917, 100, \dodoi{10.3847/1538-4357/ac0a80}

\bibitem[{McClintock {et~al.}(2014)McClintock, Norton, \& Li}]{McClintock2014}
McClintock, B.~H., Norton, A.~A., \& Li, J. 2014, The Astrophysical Journal,
  797, 130, \dodoi{10.1088/0004-637x/797/2/130}

\bibitem[{Mu{\~{n}}oz-Jaramillo {et~al.}(2013)Mu{\~{n}}oz-Jaramillo,
  Dasi-Espuig, Balmaceda, \& DeLuca}]{Andres2013}
Mu{\~{n}}oz-Jaramillo, A., Dasi-Espuig, M., Balmaceda, L.~A., \& DeLuca, E.~E.
  2013, 767, L25, \dodoi{10.1088/2041-8205/767/2/l25}

\bibitem[{Mu{\~{n}}oz-Jaramillo {et~al.}(2021)Mu{\~{n}}oz-Jaramillo, Navarrete,
  \& Campusano}]{Andres2021}
Mu{\~{n}}oz-Jaramillo, A., Navarrete, B., \& Campusano, L.~E. 2021, 920, 31,
  \dodoi{10.3847/1538-4357/ac133b}

\bibitem[{Mu{\~{n}}oz-Jaramillo {et~al.}(2015)Mu{\~{n}}oz-Jaramillo, Senkpeil,
  Windmueller, Amouzou, Longcope, Tlatov, Nagovitsyn, Pevtsov, Chapman,
  Cookson, Yeates, Watson, Balmaceda, DeLuca, \& Martens}]{Andres2015}
Mu{\~{n}}oz-Jaramillo, A., Senkpeil, R.~R., Windmueller, J.~C., {et~al.} 2015,
  800, 48, \dodoi{10.1088/0004-637x/800/1/48}

\bibitem[{Muñoz-Jaramillo {et~al.}(2010)Muñoz-Jaramillo, Nandy, Martens, \&
  Yeates}]{Munoz2010}
Muñoz-Jaramillo, A., Nandy, D., Martens, P. C.~H., \& Yeates, A.~R. 2010, 720,
  L20, \dodoi{10.1088/2041-8205/720/1/L20}

\bibitem[{{Nagy} {et~al.}(2019){Nagy}, {Lemerle}, \& {Charbonneau}}]{Nagy2019}
{Nagy}, M., {Lemerle}, A., \& {Charbonneau}, P. 2019, Advances in Space
  Research, 63, 1425, \dodoi{10.1016/j.asr.2018.12.018}

\bibitem[{{Nagy} {et~al.}(2017){Nagy}, {Lemerle}, {Labonville}, {Petrovay}, \&
  {Charbonneau}}]{Nagy2017}
{Nagy}, M., {Lemerle}, A., {Labonville}, F., {Petrovay}, K., \& {Charbonneau},
  P. 2017, \solphys, 292, 167, \dodoi{10.1007/s11207-017-1194-0}

\bibitem[{{Nagy} {et~al.}(2020){Nagy}, {Petrovay}, {Lemerle}, \&
  {Charbonneau}}]{Nagy2020}
{Nagy}, M., {Petrovay}, K., {Lemerle}, A., \& {Charbonneau}, P. 2020, Journal
  of Space Weather and Space Climate, 10, 46, \dodoi{10.1051/swsc/2020051}

\bibitem[{{Nandy}(2002)}]{Nandy2002}
{Nandy}, D. 2002, \apss, 282, 209, \dodoi{10.1023/A:1021632522168}

\bibitem[{{Nandy}(2004)}]{Nandy2004}
---. 2004, \solphys, 224, 161, \dodoi{10.1007/s11207-005-4990-x}

\bibitem[{Nandy(2006)}]{Nandy2006}
Nandy, D. 2006, Journal of Geophysical Research: Space Physics, 111,
  \dodoi{https://doi.org/10.1029/2006JA011882}

\bibitem[{{Nandy}(2021)}]{Nandy2021}
{Nandy}, D. 2021, \solphys, 296, 54, \dodoi{10.1007/s11207-021-01797-2}

\bibitem[{{Nandy} {et~al.}(2018){Nandy}, {Bhowmik}, {Yeates}, {Panda},
  {Tarafder}, \& {Dash}}]{Nandy2018}
{Nandy}, D., {Bhowmik}, P., {Yeates}, A.~R., {et~al.} 2018, \apj, 853, 72,
  \dodoi{10.3847/1538-4357/aaa1eb}

\bibitem[{Nandy \& Choudhuri(2001)}]{Nandy2001}
Nandy, D., \& Choudhuri, A.~R. 2001, The Astrophysical Journal, 551, 576,
  \dodoi{10.1086/320057}

\bibitem[{Nandy \& Martens(2007)}]{Nandy2007}
Nandy, D., \& Martens, P. 2007, Advances in Space Research, 40, 891,
  \dodoi{10.1016/j.asr.2007.01.079}

\bibitem[{Nandy {et~al.}(2023)Nandy, Banerjee, Bhowmik, Brun, Cameron, Gibson,
  Hanasoge, Harra, Hassler, Jain, Jiang, Jouve, Mackay, Mahajan, Mandrini,
  Owens, Pal, Pinto, Saha, Sun, Tripathi, \& Usoskin}]{Nandy2023arxiv}
Nandy, D., Banerjee, D., Bhowmik, P., {et~al.} 2023, Exploring the Solar Poles:
  The Last Great Frontier of the Sun,  arXiv, \dodoi{10.48550/ARXIV.2301.00010}

\bibitem[{{Parker}(1955)}]{Parker1955}
{Parker}, E.~N. 1955, \apj, 122, 293, \dodoi{10.1086/146087}

\bibitem[{{Petrovay} {et~al.}(2020){Petrovay}, {Nagy}, \&
  {Yeates}}]{Petrovay2020}
{Petrovay}, K., {Nagy}, M., \& {Yeates}, A.~R. 2020, Journal of Space Weather
  and Space Climate, 10, 50, \dodoi{10.1051/swsc/2020050}

\bibitem[{{Piddington}(1972)}]{Piddington1972}
{Piddington}, J.~H. 1972, \solphys, 22, 3, \dodoi{10.1007/BF00145457}

\bibitem[{{Schmidt}(1968)}]{Schmidt1968}
{Schmidt}, H.~U. 1968, in Structure and Development of Solar Active Regions,
  ed. K.~O. {Kiepenheuer}, Vol.~35, 95

\bibitem[{{Schou} {et~al.}(1998){Schou}, {Antia}, {Basu}, {Bogart}, {Bush},
  {Chitre}, {Christensen-Dalsgaard}, {Di Mauro}, {Dziembowski}, {Eff-Darwich},
  {Gough}, {Haber}, {Hoeksema}, {Howe}, {Korzennik}, {Kosovichev}, {Larsen},
  {Pijpers}, {Scherrer}, {Sekii}, {Tarbell}, {Title}, {Thompson}, \&
  {Toomre}}]{Schou1998}
{Schou}, J., {Antia}, H.~M., {Basu}, S., {et~al.} 1998, \apj, 505, 390,
  \dodoi{10.1086/306146}

\bibitem[{{Schrijver} {et~al.}(2015){Schrijver}, {Kauristie}, {Aylward},
  {Denardini}, {Gibson}, {Glover}, {Gopalswamy}, {Grande}, {Hapgood},
  {Heynderickx}, {Jakowski}, {Kalegaev}, {Lapenta}, {Linker}, {Liu},
  {Mandrini}, {Mann}, {Nagatsuma}, {Nandy}, {Obara}, {Paul O'Brien}, {Onsager},
  {Opgenoorth}, {Terkildsen}, {Valladares}, \& {Vilmer}}]{Schriver2015}
{Schrijver}, C.~J., {Kauristie}, K., {Aylward}, A.~D., {et~al.} 2015, Advances
  in Space Research, 55, 2745, \dodoi{10.1016/j.asr.2015.03.023}

\bibitem[{{Schwabe}(1844)}]{Schwabe1844}
{Schwabe}, H. 1844, Astronomische Nachrichten, 21, 233,
  \dodoi{10.1002/asna.18440211505}

\bibitem[{{Snodgrass}(1983)}]{Snodgrass1983}
{Snodgrass}, H.~B. 1983, \apj, 270, 288, \dodoi{10.1086/161121}

\bibitem[{Sokoloff {et~al.}(2015)Sokoloff, Khlystova, \&
  Abramenko}]{Sokoloff2015}
Sokoloff, D., Khlystova, A., \& Abramenko, V. 2015, \mnras, 451, 1522,
  \dodoi{10.1093/mnras/stv1036}

\bibitem[{Stenflo \& Kosovichev(2012)}]{Stenflo2012}
Stenflo, J.~O., \& Kosovichev, A.~G. 2012, The Astrophysical Journal, 745, 129,
  \dodoi{10.1088/0004-637x/745/2/129}

\bibitem[{Upton \& Hathaway(2014)}]{Upton2014}
Upton, L., \& Hathaway, D.~H. 2014, 792, 142,
  \dodoi{10.1088/0004-637x/792/2/142}

\bibitem[{van Ballegooijen {et~al.}(1998)van Ballegooijen, Cartledge, \&
  Priest}]{vanBallegooijen1998}
van Ballegooijen, A.~A., Cartledge, N.~P., \& Priest, E.~R. 1998, \apj, 501,
  866, \dodoi{10.1086/305823}

\bibitem[{Waldmeier(1935)}]{Waldmeier1935}
Waldmeier, M. 1935, Astronomische Mitteilungen der Eidgen{\"o}ssischen
  Sternwarte Zurich, 14, 105

\bibitem[{{Wang} {et~al.}(2002){Wang}, {Lean}, \& {Sheeley}}]{Wang2002ApJ}
{Wang}, Y.~M., {Lean}, J., \& {Sheeley}, N.~R., J. 2002, \apjl, 577, L53,
  \dodoi{10.1086/344196}

\bibitem[{{Wang} {et~al.}(1991){Wang}, {Sheeley}, \& {Nash}}]{Wang1991}
{Wang}, Y.~M., {Sheeley}, N.~R., J., \& {Nash}, A.~G. 1991, \apj, 383, 431,
  \dodoi{10.1086/170800}

\bibitem[{{Weber} {et~al.}(2013){Weber}, {Fan}, \& {Miesch}}]{Weber2013}
{Weber}, M.~A., {Fan}, Y., \& {Miesch}, M.~S. 2013, \solphys, 287, 239,
  \dodoi{10.1007/s11207-012-0093-7}

\bibitem[{{Yeates}(2020)}]{Yeates2020}
{Yeates}, A.~R. 2020, \solphys, 295, 119, \dodoi{10.1007/s11207-020-01688-y}

\bibitem[{{Yeates} {et~al.}(2015){Yeates}, {Baker}, \& {van
  Driel-Gesztelyi}}]{Yeates2015}
{Yeates}, A.~R., {Baker}, D., \& {van Driel-Gesztelyi}, L. 2015, \solphys, 290,
  3189, \dodoi{10.1007/s11207-015-0660-9}

\bibitem[{{Yeates} {et~al.}(2008){Yeates}, {Nandy}, \& {Mackay}}]{Yeates2008}
{Yeates}, A.~R., {Nandy}, D., \& {Mackay}, D.~H. 2008, \apj, 673, 544,
  \dodoi{10.1086/524352}

\bibitem[{{Zhukova} {et~al.}(2020){Zhukova}, {Khlystova}, {Abramenko}, \&
  {Sokoloff}}]{Zhukova2020}
{Zhukova}, A., {Khlystova}, A., {Abramenko}, V., \& {Sokoloff}, D. 2020,
  \solphys, 295, 165, \dodoi{10.1007/s11207-020-01734-9}

\end{thebibliography}
%\pagebreak

\begin{table}[ht]
\caption{A list of quantitative results: Anomalous active regions (AARs) coming from distinct distributions}

\label{T-complex}
\begin{center}
\begin{tabular}{ |p{1cm}|p{6cm}|p{5cm}|p{4cm}| }
\hline
\textbf{Sl No.} & \textbf{Type of AAR distribution} & \textbf{Decrease in Dipole Moment} & \textbf{Delay in time reversal} \\
\hline
1) & 5 $\%$ AH-J all over the cycle (5$\%$ flux) & 13.7 $\%$ & 2.61 $\%$ \\
\hline
2) & 10 $\%$ AH-J all over the cycle (10$\%$ flux) & 48.17 $\%$ & 7.85 $\%$ \\

\hline
\hline

3) & 5 $\%$ AH-J appearing more at the initial phase & 3.99 $\%$ &
 3.14 $\%$ \\
\hline
4) & 5 $\%$ AH-J appearing more at the middle phase & 11.04 $\%$ & 5.75 $\%$ \\
\hline
5) & 5 $\%$ AH-J appearing more at the end phase & 26.17 $\%$ & 
1.57 $\%$ \\
\hline
6) & 5 $\%$ AH-J appearing more at the upper latitude & 3.73 $\%$ & 3.14 $\%$ \\
\hline
7) & 5 $\%$ AH-J appearing more at the middle latitude & 9.75 $\%$ & 3.14 $\%$ \\
\hline
8) & 5 $\%$ AH-J appearing more near the equator & 40.64 $\%$ & 4.71 $\%$ \\

\hline
\hline
\end{tabular}
\end{center}\label{table:1}
\end{table}

%\pagebreak
%% Figures 
%

%% This command is needed to show the entire author+affiliation list when
%% the collaboration and author truncation commands are used.  It has to
%% go at the end of the manuscript.
%\allauthors

%% Include this line if you are using the \added, \replaced, \deleted
%% commands to see a summary list of all changes at the end of the article.
%\listofchanges

\end{document}